\documentclass[aps,pra,twocolumn,showpacs,showkeys,amsmath,superscriptaddress,floatfix]{revtex4-1}

\usepackage{amsmath}
\usepackage{makeidx}
\usepackage{amsfonts}
\usepackage[utf8]{inputenc}
\usepackage[usenames,dvipsnames]{pstricks}
\usepackage{subfigure}
\usepackage{epsfig}
\usepackage{mathtools}
\usepackage{grffile}
\usepackage{pst-grad} 
\usepackage{pst-plot} 
  \usepackage[colorlinks,hyperindex]{hyperref}
\hypersetup
{
	colorlinks,%
	citecolor=black,%
	linkcolor=black,%
	urlcolor=black,%
}

\renewcommand{\vec}[1]{\mathbf{#1}}
\makeindex
\begin{document}
	
	\title{Persistent oscillations of the order parameter and interaction quench phase diagram for a confined Bardeen-Cooper-Schrieffer Fermi gas}
	
	\author{S.~Hannibal}
	\affiliation{Institut f\"ur Festk\"orpertheorie, Westf\"alische
		Wilhelms-Universit\"at M\"unster, 48149 M\"unster,
		Germany}
	 
	\author{P.~Kettmann}
	\affiliation{Institut f\"ur Festk\"orpertheorie, Westf\"alische
		Wilhelms-Universit\"at M\"unster, 48149 M\"unster,
		Germany}
	
	\author{M.~D.~Croitoru}
	\affiliation{Theoretische Physik III, Universit\"at Bayreuth, 95440
		Bayreuth,
		Germany}
	
	
	\author{V.~M.~Axt}
	\affiliation{Theoretische Physik III, Universit\"at Bayreuth, 95440
		Bayreuth,
		Germany}
	
	\author{T.~Kuhn}
	\affiliation{Institut f\"ur Festk\"orpertheorie, Westf\"alische
		Wilhelms-Universit\"at M\"unster, 48149 M\"unster,
		Germany}
	
	\date{\today}
	
	\begin{abstract}
	We present a numerical study of the interaction quench dynamics in a superfluid ultracold Fermi gas confined in a three-dimensional cigar-shaped harmonic trap. In the present paper we investigate the amplitude mode of the superfluid order parameter after interaction quenches which start deep in the BCS phase and end in the BCS-BEC crossover regime. To this end, we exploit the Bogoliubov-de Gennes formalism which takes the confinement potential explicitly into account and provides a microscopic fully coherent description of the system. We find an anharmonic nonlinear oscillation of the modulus of the superfluid order parameter, i.e., of the Higgs mode. This oscillation persists for large times with only a small amplitude modulation being visible. We connect the frequency and the mean value of this oscillation with the breaking of Cooper pairs in the superfluid phase. Additionally, we demonstrate that the occurrence of this persistent oscillation is connected to the onset of chaotic dynamics in our model. Finally, we calculate an interaction quench phase diagram of the Higgs mode for quenches on the BCS side of the BCS-BEC crossover and discuss its properties as a function of the aspect ratio of the cigar-shaped trap.
	\end{abstract}
	
	
	
	\maketitle
	
	\section{Introduction}\label{sec:Introduction}

	Exploring collective excitations of quantum many-body systems provides key insights in order to gain a deeper understanding of the quantum nature of such systems. The nature of such collective excitations changes dramatically in the context of a spontaneously broken continuous symmetry like the U(1) symmetry present in solid state physics, i.e., in superconducting and superfluid systems, as well as in the standard model of particle physics \cite{Chatrchyan2012Observation,Higgs1964Broken}. The spontaneous symmetry breaking (SSB) of the U(1) symmetry leads to the emergence of two fundamental collective modes: the massive (Higgs) amplitude mode and the massless (Goldstone) phase mode \cite{Nambu1960Quasi,Goldstone1961Field,Higgs1964Broken}. These collective modes are important probes of the many-body system.
	
	Ultracold Fermi gases provide an ideal testbed to investigate the collective modes of superfluid systems, because  
	they ensure a very high controllability of most relevant system parameters \cite{Bloch2008Many,Giorgini2008Theory}. Using a Feshbach resonance \cite{Chin2010Feshbach} the interactions in an ultracold Fermi gas can be tuned. A weak attractive interaction leads to a Bardeen Cooper Schrieffer (BCS) superfluid state and in the case of a strong attractive interaction the fermionic constituents form dimers which condensate in a Bose-Einstein condensate (BEC). Those two regimes are connected by the continuous BCS-BEC crossover \cite{Randeria2011BCS}. With this control of the interaction an excitation of the system may be evoked by changing the interaction strength instantaneously, i.e., on a time scale much faster than the characteristic time scales of the dynamics. Such an interaction quench can be exploited to obtain direct access to the collective modes of the system. The experimental implementation, however, remains challenging. It has been proposed to be done either by a radio-frequency (RF) flip of one spin state of the superfluid to another spin state  (cf. \cite{Froehlich2011Radio}), which exhibits a different interaction strength due to the Feshbach resonance, or an optical control of a Feshbach resonance \cite{Clark2015Quantum}.

	Evidence of the Higgs mode has been found in various condensed matter and quantum gas systems, namely in charge density wave compounds \cite{Sooryakumar1980Raman,Sooryakumar1981Raman,Littlewood1981Gauge,Littlewood1982Amplitude,Soto-Garrido2017Higgs,Grasset2018Higgs} and materials \cite{Demsar1999Single,Schaefer2014Collective,Mertelj2013Incoherent}, in NbN, Nb$_{1-x}$Ti$_x$N, and NbSe$_2$ superconductors \cite{Matsunaga2012Nonequilibrium,Matsunaga2013Higgsa,Matsunaga2014Light,Sherman2015Higgs,Matsunaga2017Polarization,Measson2014Amplitude}, in quantum antiferromagnets \cite{Rueegg2008Quantum,Merchant2014Quantum,Kuroe2012Longitudinal}, in the superfluid $^3$He \cite{Avenel1980Field,Collett2013Zeeman}, and in an ultracold bosonic gas in a 2D optical lattice \cite{Bissbort2011Detecting,Endres2012Higgs}. However, a time-resolved detection of the non-adiabatic regime in the Higgs mode, as has been predicted \cite{Papenkort2007Coherent, Papenkort2008Coherent} and achieved in a superconducting samples \cite{Matsunaga2013Higgsa,Matsunaga2014Light,Matsunaga2012Nonequilibrium}, has not been reported so far for ultracold Fermi gases. Nevertheless, there are numerous theoretical studies devoted to the Higgs mode in ultracold Fermi gases and various superconducting systems \cite{Volkov1973Collisionless,Yuzbashyan2006Dynamical,Barankov2006Synchronization,Yuzbashyan2006Relaxationa, Barankov2006Synchronization, Barankov2004Collective,Dzero2007Spectroscopic,Bruun1999BCS, Zachmann2013Ultrafast, Kettmann2017Spectral, Scott2012Rapid, Hannibal2015Quench, Hannibal2018Dynamical, Bruun2014Long, Yuzbashyan2015Quantum,Chou2017Twisting,Cea2014Optical,Fischer2018Short,Sentef2016Theory,Moor2017Amplitude,Murotani2017Theory,Fulde1964Superconductivity,Larkin1972Influence,Croitoru2012In,Devreese2011Controlling}.
	
	The topics covered in these studies include: (i) a damped oscillation of the order parameter \cite{Volkov1973Collisionless}, (ii) the dynamical vanishing of the order parameter \cite{Yuzbashyan2006Dynamical,Barankov2006Synchronization}, (iii) relaxation and persistent oscillations of the order parameter \cite{Yuzbashyan2006Relaxationa, Barankov2006Synchronization, Barankov2004Collective}, (iv) spectroscopic signatures of nonequilibrium pairing in fermionic condensates \cite{Dzero2007Spectroscopic}, (v) collective excitations in confined systems \cite{Bruun1999BCS, Zachmann2013Ultrafast, Kettmann2017Spectral, Scott2012Rapid, Hannibal2015Quench, Hannibal2018Dynamical} and in a 2D system \cite{Bruun2014Long}, (vi) an entire quantum quench phase diagram of the superfluid condensate \cite{Yuzbashyan2015Quantum,Chou2017Twisting}, (vii) collective modes in strongly disordered superconductors \cite{Cea2014Optical}, (viii) the Higgs mechanism and the stability of the BCS mean-field theory in the weak coupling limit \cite{Fischer2018Short}, (ix) the influence of phonon-mediated interactions \cite{Sentef2016Theory}, (x) the Higgs mode in a moving condensate \cite{Moor2017Amplitude}, (xi) collective modes in a two-band BCS superconductor \cite{Murotani2017Theory}, and (xii) the modes in the FFLO state \cite{Fulde1964Superconductivity,Larkin1972Influence,Croitoru2012In,Devreese2011Controlling}.
	
	The general picture which emerges from the theory work is that the perturbation of a 3D system of interacting Fermions leads to a variety of dynamical phases with properties quite distinct from the equilibrium ones. A classification of the nonequilibrium behavior arising from different excitations of the BCS state has been established  for a three-dimensional homogeneous Fermi gas by calculating a quantum quench phase diagram, which has revealed three distinct phases \cite{Yuzbashyan2015Quantum}. In ``phase I'' the superfluid order parameter vanishes dynamically \cite{Yuzbashyan2015Quantum, Barankov2006Synchronization, Yuzbashyan2006Dynamical}, in ``phase II'' a damped oscillation with an $t^{-1/2}$ behavior was found \cite{Volkov1973Collisionless,Yuzbashyan2015Quantum}, and ``phase III'' is characterized by a persistent oscillation with a constant amplitude \cite{Yuzbashyan2015Quantum, Yuzbashyan2006Relaxationa, Barankov2004Collective, Barankov2006Synchronization}. Furthermore, studies in nano-structured superconducting BCS systems where quantum confinement may strongly influence the superconducting pairing \cite{Shanenko2006Size,Croitoru2007Dependence,Shanenko2008Superconducting,Chen2009Superconducting,Croitoru2009Superconducting,Shanenko2010Giant,Croitoru2012Cooper} show an alteration of the damping in phase II depending on the geometry \cite{Zachmann2013Ultrafast,Kettmann2017Spectral} while in an inhomogeneous Fermi gas with a harmonic confinement in one dimension and a box potential in the other two dimensions the homogeneous behavior has been confirmed \cite{Scott2012Rapid}.
	
	In our previous articles, we have studied the dynamical phases I and II for a Fermi gas in a three dimensional harmonic trap and we have found a confinement-induced fragmentation of the Higgs mode in phase II \cite{Hannibal2015Quench} and an alteration of the dynamical vanishing in phase I due to the confinement \cite{Hannibal2018Dynamical}. In this paper, we continue investigating the impact of the confinement potential on the Higgs mode of a superfluid order parameter and present a numerical study of the dynamical phase III as well as a quench phase diagram in a cigar-shaped ultracold Fermi gas with a harmonic confinement.

	The paper is organized as follows: Section~\ref{sec:model} introduces the model system and we will briefly sketch the used Bogoliubov-de Gennes formalism. In section~\ref{sec:persistent} we will analyze the persistent dynamics for interaction quenches starting deep in the BCS phase and ending in the BCS-BEC crossover regime. Finally, in Sec.~\ref{sec:phasediagramm} we discuss the interaction quench phase diagram of an inhomogeneous Fermi gas on the BCS side of the BCS-BEC crossover. Here, we focus on the dependence on the aspect ratio of our cigar-shaped harmonic trap. In Sec.~\ref{sec:summary} we provide a brief summary of the results obtained in the paper.

	\section{Theoretical Model}\label{sec:model}
	
	In our model we consider an ultracold Fermi gas composed of fermionic $^6$Li atoms in two different internal spin states (labeled by $\uparrow$ and $\downarrow$) with a balanced population, i.e., $N_\uparrow = N_\downarrow = N/2$. We choose a cigar-shaped harmonic confinement potential with an aspect ratio $r = f_\perp / f_\parallel \gg 1$ where $f_\perp$ ($f_\parallel$) is the radial (longitudinal) trap frequency. Further, we assume an attractive interaction between Fermions of different spins with zero range such that its strength can be characterized by a single parameter, the so-called scattering length $a$. This is commonly done for s-wave scattering processes in ultracold gases and results in an effective contact potential $V_\text{int} =  (4\pi \hbar a / m_0) \,\delta(\vec{r}) =  g \, \delta(\vec{r}) $, where $m_0$ is the mass of $^6$Li. Our treatment here is restricted to negative scattering lengths, i.e., to the BCS side of the BCS--BEC crossover. We aim at describing the superfluid phase which evolves in such an ultracold Fermi gas by means of the Bogoliubov-de Gennes (BdG) formalism.
	
	In the following, we will sketch the formalism shortly, for a detailed discussion we refer the reader to Refs. \cite{Hannibal2015Quench, Datta1999Can, DeGennes1989Superconductivity}. We start from the usual Hamiltonian for a contact interaction written in terms of the field operators. Then, we introduce the superfluid order parameter by applying a BCS-like mean-field approximation which yields the BdG Hamiltonian
	\begin{align}\label{eq:Hamiltonian}
		H_\text{BdG} = \sum_\sigma &\int d^3r\,  \Psi_\sigma^\dagger (\vec{r})  H_0 \Psi_\sigma(\vec{r}) \notag \\  + &\int d^3r \, \Delta^*(\vec{r}) \Psi_\downarrow(\vec{r})  \Psi_\uparrow(\vec{r}) + h.c. \, ,
	\end{align}
	where $H_0$ is the one-particle Hamiltonian including the kinetic energy of the bare atoms and the confinement potential. Then, the superfluid order parameter is defined by
	\begin{equation} \label{eq:order}
		\Delta(\vec{r}) = g \left< \Psi_\downarrow(\vec{r})\Psi_\uparrow(\vec{r})\right>.
	\end{equation}
	Here, $\Psi_\sigma(\vec{r})$ is a field operator annihilating an atom with spin $\sigma$ at position $\vec{r}$. In order to diagonalize the Hamiltonian in Eq.~\eqref{eq:Hamiltonian} and obtain the ground-state order parameter of the system we introduce Bogoliubov's quasiparticles:
	\begin{align}\label{eq:Bogolon}
	  \gamma^\dagger_{na} = \int u_n(\vec{r}) \Psi_\uparrow^\dagger(\vec{r}) + v_n(\vec{r}) \Psi_\downarrow(\vec{r})\, d^3 r  \\ 
	  \gamma^\dagger_{nb} = \int u_n(\vec{r}) \Psi_\downarrow^\dagger(\vec{r}) - v_n(\vec{r}) \Psi_\uparrow(\vec{r})\, d^3 r.
	\end{align}
	The indices $a/b$ result from the two spin species and $u_n(\vec{r})$ and $v_n(\vec{r})$ are the eigenfunctions of the corresponding eigenvalue problem of the Hamiltonian in Eq.~\eqref{eq:Hamiltonian}, the so-called BdG equation. Exploiting Anderson's approximation, which only considers pairing in time reversed states \cite{Anderson1959Theory}, we obtain the BdG eigenenergies $E_n = \sqrt{\varepsilon_n^2 + \Delta_n^2}$, where $\varepsilon_n$ are the eigenenergies of the atoms in the trap and $\Delta_n = \left<n|\Delta(r)|n\right>$. Here, we use $\Delta(\textbf{r}) = \sum_n u_n(\textbf{r}) v_n(\textbf{r})$ and the eigenstates of the confinement potential $\left|m\right>$ . From the solution of the BdG equation we obtain the well-known self-consistency equations for the order parameter and the chemical potential $\mu$ where the latter is determined by a given particle number $N$. We regularize and solve the obtained self-consistency equations numerically for $T=0\,$K which yields the BdG eigenenergies and eigenfunctions. Here, the numerical cutoff is chosen such that all states with energies up to twice the Fermi energy $E_F$ are considered. By inverting Bogoliubov's transformation we obtain $\Delta(\vec{r})$ according to Eq.~\eqref{eq:order}. 
	
	In order to excite a non-equilibrium state of the system we consider an instantaneous change, i.e., a quench, of the scattering length  $a_i \rightarrow a_f$. We assume that the change of the scattering length takes place on a time scale much faster than typical system reaction times. Thus, the density $n(\vec{r}) = \left<\Psi_\uparrow^\dagger \Psi_\uparrow^{\phantom{\dagger}}\right> + \left<\Psi_\downarrow^\dagger \Psi_\downarrow^{\phantom{\dagger}}\right>$ immediately after the quench is the same as before, i.e., the system is in a non-equilibrium state with respect to the new interaction strength. In order to evaluate the subsequent dynamics it is convenient to express the ground state of the initial system in terms of the eigenfunctions corresponding to the final ground state. This yields occupations  $x_{ml} = \left<\gamma_{ma}^\dagger \gamma_{la}^{\phantom{\dagger}}\right> = \left<\gamma_{mb}^\dagger \gamma_{lb}^{\phantom{\dagger}}\right> $ and coherences $y_{ml} = \left<\gamma_{ma}^{\dagger}\gamma_{lb}^{{\dagger}}\right> = \left<\gamma_{lb}^{\phantom{\dagger}} \gamma_{ma}^{\phantom{\dagger}}\right>^*$ of the density matrix instantaneously after the quench. However, an interaction quench does only introduce excitations which are diagonal, i.e., $x_{ml} \eqqcolon x_m\,\delta_{ml}$ and $y_{ml} \eqqcolon y_m\,\delta_{ml}$. These excitations provide the initial values for our non-adiabatic dynamics.
	
	
	In the next step, we make use of Heisenberg's equation of motion and obtain coupled ordinary differential equations for the occupations and coherences.
	  We solve these equations numerically and finally invert Bogoliubov's transformation in order to obtain the superfluid order parameter  
	  \begin{align}\label{eq:Delta_Bogolon}
	      \Delta(\vec{r},t) =  g  \sum_{n} \, & 2\, v_n(\vec{r}) u_n(\vec{r})\, \left( x_n(t) - \frac{1}{2}\right) \notag \\ &+ \, u_n^2(\vec{r})\, y_n^*(t) - v_n^2(\vec{r})\, y_n(t).
	  \end{align}
	  in terms of the occupations and coherences. This formalism provides the coherent quantum mechanical dynamics of an ultracold Fermi gas confined in a cigar-shaped harmonic trap in mean-field approximation.

	  \section{Persistent oscillations in the interaction quench dynamics}\label{sec:persistent}
	  
	  In this section we will investigate interaction quenches which start in the BCS regime [$1/(k_Fa_i) < -1$] and end in the BCS--BEC crossover region [$-1 < 1/(k_Fa_f) < 0$]. Here, we follow the usual characterization of the coupling strength of the system in terms of $1/(k_Fa)$, where we obtain $k_F = \sqrt{(2m_0E_F)/ \hbar^2}$ from the Fermi energy $E_F$ by the dispersion relation of free atoms. In the following we will discuss the spatiotemporal dynamics of the order parameter $\Delta(\vec{r},t)$ of an ultracold Fermi gas after such a quench. 
	  
	  \begin{figure}[t]
		\includegraphics[width=1\columnwidth]{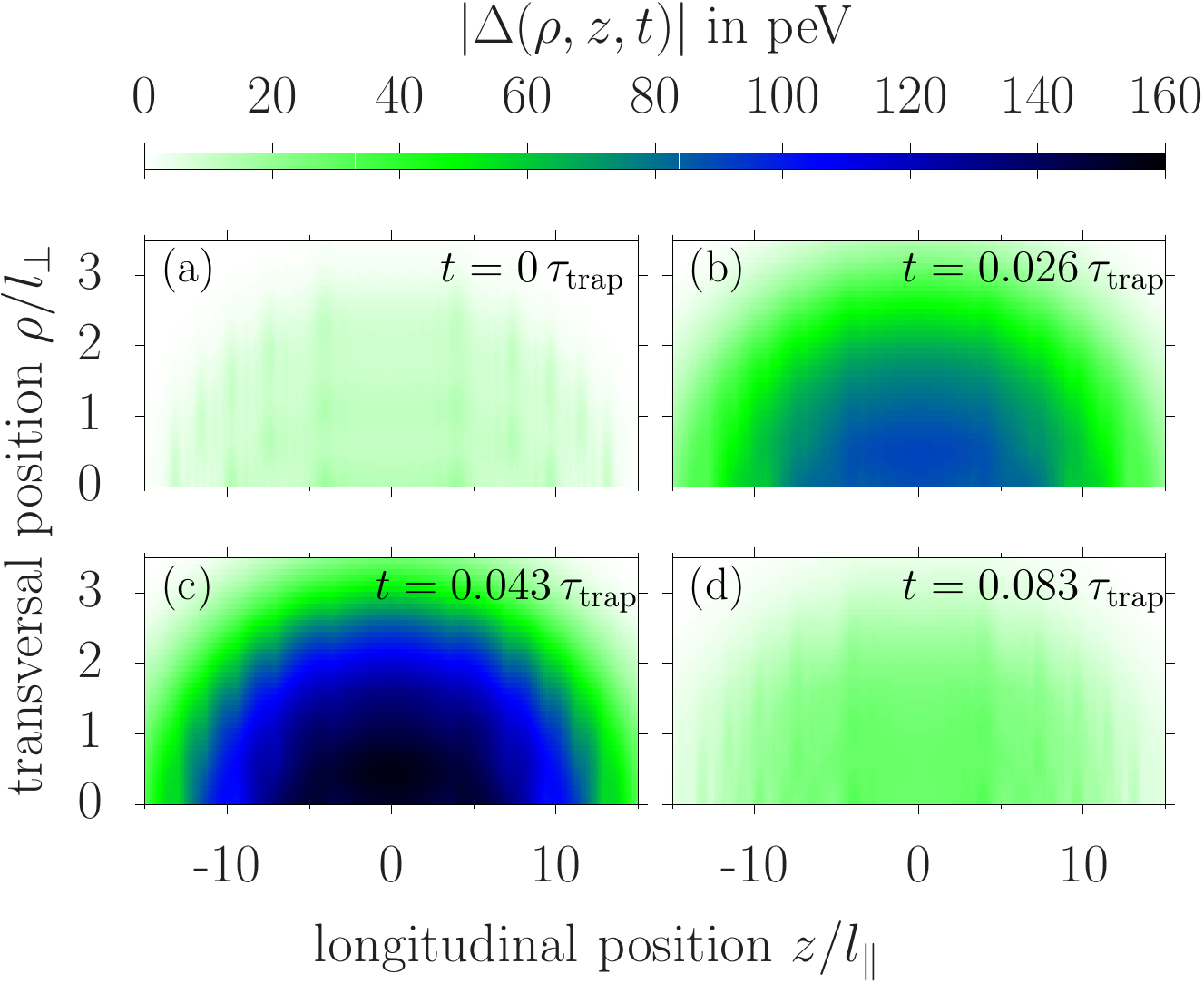}
		\caption{Spatiotemporal dynamics of the modulus of the order parameter after a sudden change of the scattering length from $1/(k_Fa_i) = -2.0$ to $1/(k_Fa_f) = -0.5$ and an aspect ratio of $r=20$. $\tau_\text{trap} = h/\delta E = 1/(2 f_\parallel) \approx 4.2\,$ms is the characteristic time scale of the trap \cite{Hannibal2015Quench}.}
		\label{fig:spatiotemporal}
	  \end{figure}
	  
	  In Fig.~\ref{fig:spatiotemporal} we show the spatiotemporal dynamics of the order parameter after an interaction quench starting from an initial coupling of $1/(k_Fa_i)=-2.0$ to a final coupling of $1/(k_fa_f) = -0.5$. We choose an aspect ratio of $r=20$ with $f_\parallel = 120\,$Hz. Furthermore, to ensure the comparability of the different systems considered in this paper we choose a constant system parameter $s = E_F / (h f_\perp) = 5.5$ for all calculations with $h$ being Plank's constant. This parameter characterizes the band structure which results from the cigar-shaped trap and the value of $s = 5.5$ is chosen such that the system size is small enough for a systematic numerical analysis and at the same time the quantum-size oscillations occurring for small systems are avoided \cite{Shanenko2012Atypical}. In such a case the results can easily be scaled to a larger system size by making use of the scaling relations which we discussed in Ref.~\cite{Hannibal2018Dynamical}. Furthermore, the parameters chosen here are exemplary for strong quenches from the BCS regime towards the BCS--BEC crossover, as we will see later from the phase diagrams in Sec.~\ref{sec:phasediagramm}.
	  
	  Figure~\ref{fig:spatiotemporal}~(a) shows the order parameter instantaneously after the quench at $t=0 \, \tau_\text{trap}$, i.e., the excitations introduced by the quench are already included. The striped distribution of $\Delta(\vec{r})$ --which is typical for the BCS regime-- corresponds to the order parameter of the initial system scaled by a factor of $a_f/a_i$. In the subsequent dynamics in Fig.~\ref{fig:spatiotemporal}~(b) we observe an increase of the order parameter which affects the whole trap and which finds its maximum at $t =0.043 \, \tau_\text{trap}$ shown in Fig.~\ref{fig:spatiotemporal}~(c). This rise in the order parameter is symmetrical with respect to the trap center and the striped structure in the initial distribution transforms into a monotonic distribution with a maximum in the center of the trap. This spatial distribution is characteristic in the BCS-BEC crossover regime and results from the order parameter in the ground state of the final system. Figure~\ref{fig:spatiotemporal}~(d) shows the succeeding minimum at $t = 0.083\,\tau_\text{trap}$. Here, again the whole trap is affected and the order parameter is slightly larger than in the initial frame while it recovers its striped structure.
	  
	  Overall, we find an oscillation that affects the whole trap while the spatial distribution of the gap oscillates between the distribution given by the ground states of the initial and of the final system, respectively. Hence, in this case the modulus of the spatially averaged order parameter correctly provides all information necessary, i.e., the frequency and amplitude of the oscillation for a detailed investigation. Hence, we introduce the spatially averaged order parameter
	  \begin{equation}
	   \overline{\Delta}(t) = \left|\frac{1}{V} \int d^3r \Delta(\vec{r},t)\right|,
	  \end{equation}
	  where we set the volume to $V = (\hbar/(m_0\pi))^{3/2} f_\perp^{-1} f_\parallel^{-1/2}$.
	  
	  \begin{figure}[t]
		\includegraphics[width=1\columnwidth]{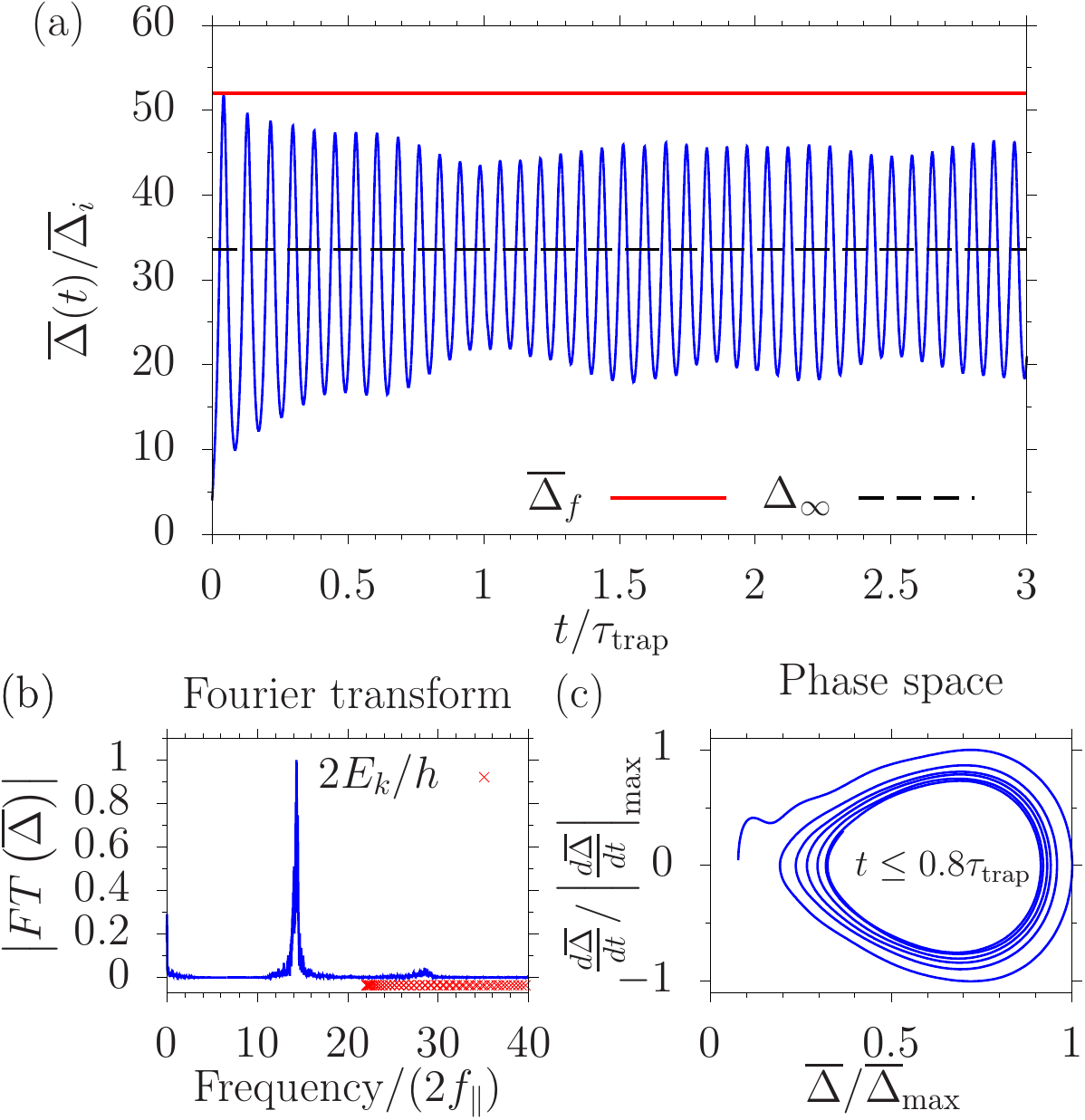}
		\caption{Amplitude mode of the spatially averaged order parameter after an interaction quench from $1/(k_Fa_i) = -2.0 $ to $1/(k_Fa_i) = -0.5$ with identical parameters as in Fig.~\ref{fig:spatiotemporal}: $r=20$, $f_\parallel = 120\,$Hz and $s=5.5$. (a) time evolution of the spatially averaged order parameter $\overline{\Delta}(t)$ normalized to the spatially averaged order parameter in the ground state of the initial system $\overline{\Delta}_i$. $\overline{\Delta}_f$ is the spatially averaged order parameter of the final system in its ground state (red line) and $\Delta_\infty$ is the mean value over the whole calculation time (black dashed line). (b) Fourier transformation of the signal in (a). $E_k$ are the Bogoliubov quasiparticle eigenenergies (red crosses). (c) phase space representation of the signal in (a) normalized to the corresponding maximum values.}
		\label{fig:deltaquer}
	  \end{figure}
	  
	  The time evolution of the modulus of the spatially averaged order parameter $\overline{\Delta}(t)$ is shown in Fig.~\ref{fig:deltaquer}~(a) normalized to the spatially averaged order parameter in the initial system $\overline{\Delta}_i$. We find an oscillation around the mean value $\Delta_\infty$ (black dashed line) which is reduced compared to the ground state value of the final system $\overline{\Delta}_f$ (solid red line). After a small initial damping the oscillation persists for several $\tau_\text{trap}$. Only a modulation of the amplitude is visible, opposed to the homogeneous case where the persistent oscillation features a constant amplitude \cite{Yuzbashyan2015Quantum,Yuzbashyan2006Relaxationa}. 
	  
	  By investigating the Fourier transform of the order parameter we observe that the persistent oscillation shows a pronounced nonlinear character: In Fig.~\ref{fig:deltaquer}~(b) we show the Fourier spectrum of $\overline{\Delta}(t)$ which shows one dominant feature at $f_\text{Higgs} = 14.2 \cdot (2 f_\parallel)$ being smaller than twice the smallest Bogoliubov eigenenergies $2 E_\text{min}/h = 22 \cdot (2f_\parallel)$ (red crosses). In contrast, in the linear case the spectrum is composed of a series of peaks given by twice the quasiparticle energies.  This leads to linear dephasing dynamics of the Higgs mode which break down after $\tau_\text{trap}$ \cite{Hannibal2015Quench}. Hence, the absence of the break down in the persistent oscillation is a consequence of the nonlinear character. Additionally, we observe a further contribution in the spectrum around $28.5 \cdot (2f_\parallel)$, which corresponds to the second harmonic of the dominant frequency, indicating that the oscillation of $\overline{\Delta}(t)$ is anharmonic. In order to illustrate this, we depict in Fig.~\ref{fig:deltaquer}~(c) the phase space of the oscillation, i.e., $d/dt (\overline{\Delta})$ vs. $\overline{\Delta}$, each being normalized to their corresponding maximum value. In this phase space we find a trajectory which is egg-shaped. This demonstrates the anharmonic character of the oscillation in accordance with what has been found in phase III in the homogeneous system \cite{Yuzbashyan2015Quantum}.
	  
	  Summarizing, we observe an anharmonic nonlinear oscillation that persists after $\tau_\text{trap}$ where the amplitude is modulated. In the next step, we will investigate the origin of the frequency of the oscillation and of the reduction of the mean value in more detail and we will demonstrate that those two are interconnected.
	  
\begin{figure}[t]
		\includegraphics[width=1\columnwidth]{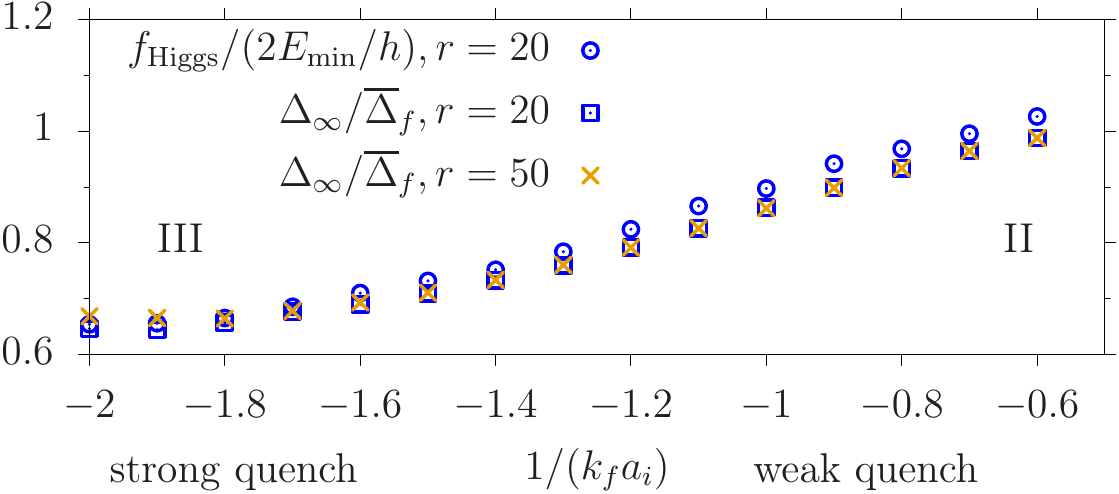}
		\caption{Frequency of the Higgs mode $f_\text{Higgs}$ normalized to twice the smallest BdG quasiparticle energy $2E_\text{min}$ (blue circles) and mean value $\Delta_\infty$ of $\overline{\Delta}(t)$ normalized to the ground state value $\overline{\Delta}_f$ (orange pluses) for various quench strengths characterized by $1/(k_Fa_i)$. The final coupling is $1/(k_Fa_f) = -0.5$ and all other parameters are identical to Fig.~\ref{fig:spatiotemporal} \& Fig.~\ref{fig:deltaquer}. The roman numbers mark ``phase II'' and ``phase III'', respectively.}
		\label{fig:fhiggs_eren}
	  \end{figure}

	  An excitation of Bogoliubov quasiparticles in the BCS phase reduces the modulus of the order parameter \cite{Ketterle2008Makingb}. This can be understood by considering that the population of Bogoliubov's quasiparticles $x_k$ corresponds to the breaking of Cooper pairs. Bearing in mind that $\left| \Delta(\vec{r})\right|^2$ is proportional to the Cooper pair density \cite{Leggett2006Quantum} it becomes apparent that such an excitation also reduces the modulus of the spatially averaged order parameter $\overline{\Delta}$ which sets the mean value of the oscillation. Furthermore, from dynamical studies in homogeneous systems it is known that the frequency of the Higgs mode, i.e., the oscillation frequencies of the modulus of the order parameter, is determined by the mean value of the modulus of the order parameter itself \cite{Papenkort2008Coherent,Papenkort2007Coherent, Yuzbashyan2015Quantum}. Hence, a reduction of $\overline{\Delta}$ due to excitations leads on the one hand to a reduction of the mean value $\Delta_\infty$ and on the other hand it implies a reduction of $f_\text{Higgs}$. Indeed, from our numerical data we do not only extract that this is the case. We also find that the reduction of $f_\text{Higgs}$ is approximately given by the reduction of $\Delta_\infty$ for all investigated quenches as we exemplary show in Fig.~\ref{fig:fhiggs_eren}. 
	  
	  Figure~\ref{fig:fhiggs_eren} shows the frequency of the Higgs mode $f_\text{Higgs}$ normalized to twice the smallest BdG quasiparticle energy $2E_\text{min}$ (blue circles) for $r=20$, and $\Delta_\infty / \overline{\Delta}_f$ for $r= 20$ (blue boxes) and $r=50$ (orange crosses) as a function of $1/(k_Fa_i)$, i.e., of the quench strength. Here, we choose the remaining parameters identical to Fig.~\ref{fig:spatiotemporal} \& Fig.~\ref{fig:deltaquer}. We choose the normalization such that both quantities ($f_\text{Higgs}$ and $\Delta_\infty$) are normalized to the value they take in the case of a linear dynamics induced by a weak quench. Both quantities in Fig.~\ref{fig:fhiggs_eren} decrease in the same nonlinear fashion when increasing the quench strength, i.e., when choosing a smaller initial coupling $1/(k_Fa_i)$. From our data we confirm that this close interconnection between $f_\text{Higgs}$ and $\Delta_\infty$ applies for all considered quenches and aspect ratios. This clearly shows a connection between both quantities. Furthermore, comparing $\Delta_\infty / \overline{\Delta}_f$ for $r=20$ and $r=50$ in Fig.~\ref{fig:fhiggs_eren} reveals that the decrease with increasing quench strength is independent of the aspect ratio $r$. In accordance to our preceding discussion, the reduction of $\Delta_\infty / \overline{\Delta}_f$ is controlled by the initial quasiparticle occupations introduced at the time of the quench, which is confirmed by the numerical data. These initial quasiparticle occupations only depend on the quench strength but not on the aspect ratio $r$ in the case of a constant system parameter $s$.
	  
	  
	  Overall, we have demonstrated that the decrease of $f_\text{Higgs}$ and $\Delta_\infty$ with increasing quench strength in the inhomogeneous system is caused by the breaking of Cooper pairs due to the creation of quasiparticle occupations. These excitations are introduced by the interaction quench and their occupation increases with the quench strength \cite{Hannibal2018Dynamical} which leads to the observed reduction of $f_\text{Higgs}$ and $\Delta_\infty$. In the next step, we will analyze the nonlinear nature of the oscillation further and we will show that our model system shows a chaotic signature when entering phase III of the quench dynamics.
	  
	  In order to test whether the dynamics of the system is chaotic we introduce a small perturbation and calculate the difference between the perturbed and unperturbed trajectory in phase space, which is in this case defined by the occupations and coherences. To this end, we add a randomized perturbation $\eta_k^{x/y}$ to the unperturbed initial values and we obtain the perturbed initial values ${x'_k} {|_{t=0}} = {x_k} {|_{t=0}} + \eta_k^x$ and ${y'_k} {|_{t=0}}= {y_k} {|_{t=0}} + \eta_k^y$. In the next step, we calculate the resulting trajectory $z' \coloneqq z'(x'_1,\dots,x'_{N_c},y'_1,\dots,y'_{N_c})$ together with the unperturbed trajectory $z \coloneqq z(x_1,\dots,x_{N_c},y_1,\dots,y_{N_c})$. Here, $N_c$ is the number of states in the calculation set by the numerical cutoff. In order to measure the difference between these two trajectories in phase space we use the norm \cite{Tabor1989Chaos}
	  \begin{equation}
	   \left\|z\right\| = \frac{1}{N_c}\sqrt{\sum_k \left(\left|x_k\right|^2 + \left|y_k\right|^2\right)}.
	  \end{equation}
	  For our numerical calculation, we choose all perturbations $\eta_k^{x/y}$ to be within the same order of magnitude and we ensure that the initial relative distance in phase space is small, i.e., $\left\|z'-z\right\| / \left\|z\right\|  \lesssim 10^{-8}$. With this choice we ensure that the numerical errors per time step in our calculations are kept at least five orders of magnitude smaller than the perturbation which provides reliable data.
	  
	  \begin{figure}[t]
		\includegraphics[width=1\columnwidth]{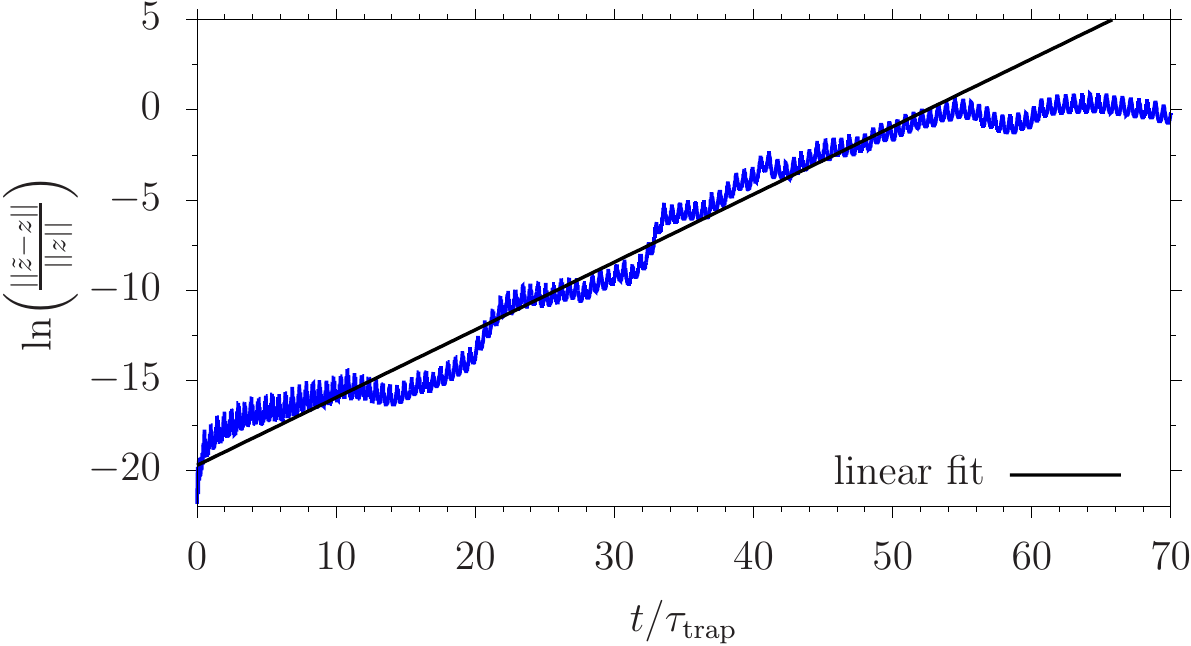}
		\caption{Temporal evolution of the logarithm of the relative difference in phase space ln$(\left\|z'-z\right\| / \left\| z \right\|)$ between the two trajectories $z$ and $z'$. The black line shows a linear fit for $t < 51\, \tau_\text{trap}$ from which we extract the largest Lyapunov exponent $x_{Ly}^{(1)}$. The parameters are identical to Fig.~\ref{fig:spatiotemporal} \& Fig.~\ref{fig:deltaquer}.}
		\label{fig:chaos}
	  \end{figure}
	  
	  In Fig.~\ref{fig:chaos} we illustrate the temporal evolution of the logarithm of the relative difference in phase space ln$(\left\|z'-z\right\| / \left\| z \right\|)$ between the two trajectories. We find that ln$(\left\|z'-z\right\| / \left\| z \right\|)$ oscillates with an underlying linear increase for $t < 55\, \tau_\text{trap}$ which saturates for larger times. The saturation sets in at ln$(\left\|z'-z\right\| / \left\| z \right\|) \approx 1 $ which corresponds to a situation where the difference in phase space is equal to the norm of the trajectory itself. Such a behavior is commonly considered as the signature of systems which exhibit chaos. In this case the exponent of the exponential increase of $\left\|z'-z\right\| / \left\| z \right\|$ is the so-called first Lyapunov exponent $x_{Ly}^{(1)}$ which determines the behavior of the systems dynamics. We extract the Lyapunov exponent from the linear fit shown in Fig.~\ref{fig:chaos} as $x_{Ly}^{(1)} \approx 0.38 / \tau_\text{trap}$. Overall, this clearly shows the chaotic signature of the dynamics.
	  
	  \begin{figure*}[t]
		\includegraphics[width=\textwidth]{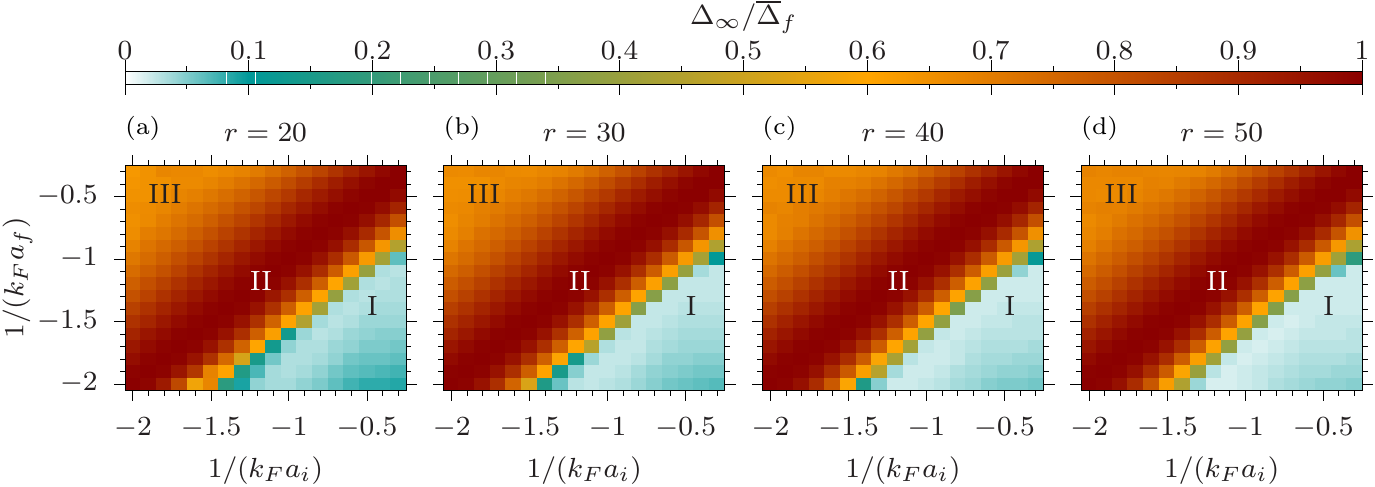}
		\caption{Interaction quench phase diagrams of $\Delta_\infty / \overline{\Delta}_f$ for aspect ratios of $r=20,30,40,50$, a longitudinal trap frequency of $f_\parallel = 120\,$Hz and a system parameter of $s = 5.5$. The dynamical phase I is marked as I and the dynamical phase III is marked as III.}
		\label{fig:phasedigram}
	  \end{figure*}
	  
	  
	  Summarizing, we have analyzed the dynamics after an interaction quench from the deep BCS regime to the BCS-BEC crossover region, i.e., the dynamical phase III. After the quench we observe an anharmonic nonlinear oscillation where both the frequency and the mean value are reduced due to the breaking of Cooper pairs which is induced by the excitation of quasiparticles at the time of the quench. Furthermore, we find that the dynamics in the BdG formalism after such a quench shows a clear signature of chaos. Having analyzed phase III in detail here and previously ``phase II'' in Ref.~\cite{Hannibal2015Quench} and ``phase I'' in Ref.~\cite{Hannibal2018Dynamical}, we will in the following complete the picture by discussing the whole quench phase diagram for an inhomogeneous system. The following analysis is similar to what has been done for a homogeneous system in the thermodynamic limes \cite{Yuzbashyan2015Quantum}.

	  \section{Quench phase diagram}\label{sec:phasediagramm}
	  
	  In this section we will consider all types of interaction quenches possible on the BCS side of the BCS-BEC crossover which yields a dynamical quench phase diagram for an inhomogeneous ultracold Fermi gas. Further, we will discuss the effect of the aspect ratio $r$ on the phase diagram. However, before we analyze this feature of the phase diagram we will locate the different phases  in the phase diagram and we will briefly recall the characteristics of the two phases which have been analyzed in detail in previous publications, i.e., phase I \cite{Hannibal2018Dynamical} and phase II \cite{Hannibal2015Quench}.

	  Figure~\ref{fig:phasedigram} shows the interaction quench phase diagram for four different aspect ratios $r=20,30,40$ and $50$. Here, we depict $\Delta_\infty/ \overline{\Delta}_f$ which shows a distinct characteristic for each phase. We choose a longitudinal trap frequency of $f_\parallel = 120\,$Hz and keep the system parameter $s=5.5$ fixed in all systems. Together with a given $r$, $1/(k_Fa_i)$ and $1/(k_Fa_f)$ this determines all remaining parameters.
	  
	  Around the diagonal in each frame of Fig.~\ref{fig:phasedigram}, i.e., for weak quenches in both directions, we find that $\Delta_\infty /\overline{\Delta} \approx 1$ which corresponds to the linear dynamics of  ``phase II''.  Here, the Higgs mode of the order parameter shows a damped oscillation which breaks down at $\tau_\text{trap}$ due to the dephasing of the linearly coupled occupations and coherences of the quasiparticle density matrix  \cite{Hannibal2015Quench}. 
	  
	  In the lower right corner of the phase diagram (marked as I) we depict quenches from the BCS-BEC crossover regime to the deep BCS regime. These interaction quenches lead to an initial exponential decay and a consecutive dynamical vanishing of the modulus of the order parameter, i.e., the evolution of a flat plateau with an almost vanishing average value. This vanishing is characterized by a small value of $\Delta_\infty / \overline{\Delta}_f$. The onset of the dynamical vanishing was found to depend crucially on the smallest quasiparticle energy of the final system $E_\text{min} = \text{Min}_k(E_k)$ which determines both the initial decay constant and the visibility of a plateau in the dynamics of $\overline{\Delta}(t)$ \cite{Hannibal2018Dynamical}. One aspect that has not been discussed previously is the increase of $\Delta_\infty / \overline{\Delta}_f$ for very large quenches, visible by the slightly darker area in the lowest right corner. In this case, the strong interaction quench excites an oscillation with the transverse confinement frequency $f_\perp$ on top of the plateau. This oscillation then prevents the modulus of the order parameter to vanish further and, thus, leads to the observed increase of $\Delta_\infty / \overline{\Delta}_f$ in the phase diagram.
	  
	  Finally, in the upper left corner of the interaction quench phase diagrams in Fig.~\ref{fig:phasedigram}~(a)-(d) we find phase III which has been discussed in Sec.~\ref{sec:persistent} and is identified by a smooth reduction of $\Delta_\infty / \overline{\Delta}_f$, as we have discussed above. We have found, that the reduction of $\Delta_\infty/ \overline{\Delta}$ is induced by the breaking of Cooper pairs at the time of the quench, which is independent of the aspect ratio $r$. With this characterization, we will now discuss the dependency of the quench dynamics on the aspect ratio by means of the obtained phase diagrams.
	  
	  Comparing the four phase diagrams in Fig.~\ref{fig:phasedigram} we observe that phase II around the diagonal and phase III in the upper left corner of the dynamical phase diagram remain essentially unchanged when changing the aspect ratio. This is expected, since we found that the breaking of Cooper pairs by quasiparticle excitations at the time of the quench leads to the observed reduction of $\Delta_\infty / \overline{\Delta}_f$. However, the amount of excitations induced only depends on the quench strength and not on the aspect ratio $r$ in the case of a constant system parameter $s$, as discussed above. 
	  
	  In contrast, investigating the lower right corner of the dynamical phase diagram, i.e., phase I, there are changes in the transition visible. In the phase diagram for $r=20$ in Fig.~\ref{fig:phasedigram}~(a) a darker blue-green area is visible at the transition between phase I and phase II for $1/(k_Fa_f) \le -1.5$. In comparison to the phase diagram for larger aspect ratios we observe that the darker blue-green stripe, which starts at the quench from $1/(k_Fa_i) = -1.0$ to $1/(k_Fa_f) = -1.6$ for $r=20$ and extends parallel to the diagonal, moves towards smaller effective interaction strength in the final system until it is not visible any more in the dynamical phase diagram for $r=50$. The observed shift results from an increase of $E_\text{min}$ with on the one hand the aspect ratio $r$ and on the other hand the interaction strength in the final system $1/(k_Fa_f)$: The feature we observe in the phase diagram, i.e., the darker blue-green area, occurs for an identical $E_\text{min}$ and quench strength for all aspect ratio $r$. Therefore, it shifts towards smaller effective interaction strength with an increasing aspect ratio $r$ in order to account for the increase of $E_\text{min}$ with $r$. This shift takes place in accordance with the dependence of the onset of the dynamical vanishing order parameter we have previously found for phase I \cite{Hannibal2018Dynamical}. In addition, an analogue effect leads to the darkening blue-green feature for the quench from $1/(k_Fa_i) = -0.3$ to $1/(k_Fa_f) = -1.0$ with an increasing aspect ratio. 
	  
	  Furthermore, we observe that the slightly darker area in the far lower right corner of the phase diagram fades with an increasing aspect ratio. We understand this in terms of the oscillation with $f_\perp$ on top of the plateau: since $f_\perp$ is increased with the aspect ratio a stronger quench is necessary in order to excite the additional oscillation with the trap frequency. Hence, this oscillations becomes less pronounced with an increasing aspect ratio.
	  
	  \begin{figure}[t]
		\includegraphics[width=1\columnwidth]{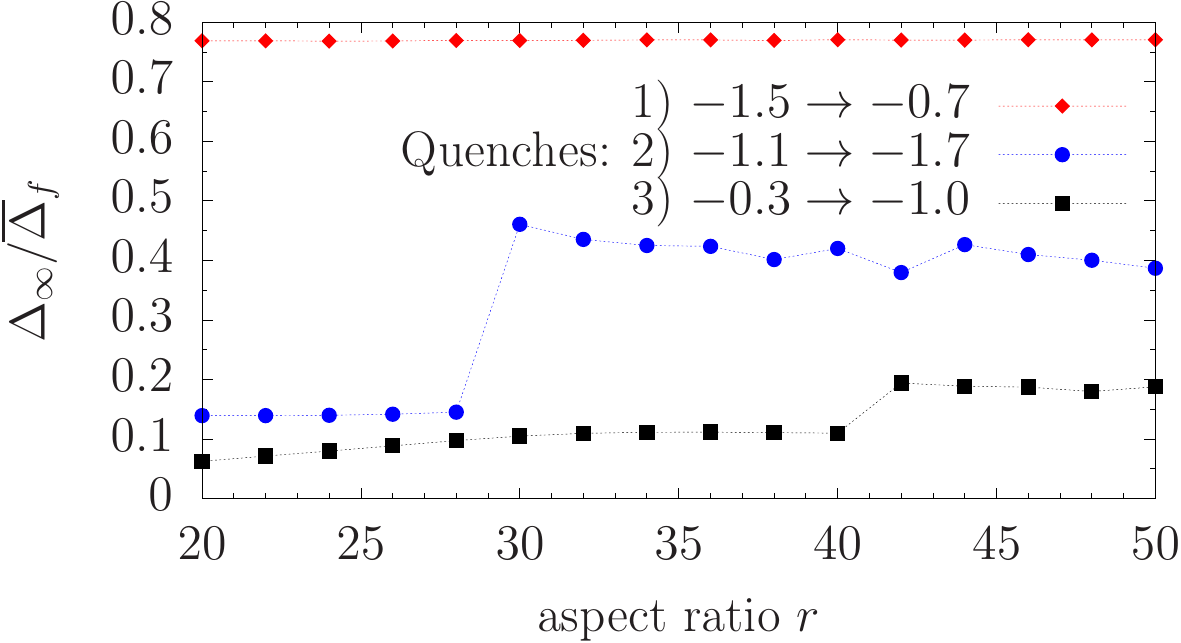}
		\caption{$\Delta_\infty / \overline{\Delta}_f$ in dependence on the aspect ratio $r$ for three exemplary quenches from the phase diagrams in Fig.~\ref{fig:phasedigram}. All parameters are identical to before and the dashed lines are a guide to the eye.}
		\label{fig:Transition_vs_r}
	  \end{figure}
	  
	  In order to highlight the dependence of the dynamical phase diagram on the aspect ratio $r$, we depict $\Delta_\infty / \overline{\Delta}_f$ vs. the aspect ratio $r$ for three exemplary quenches in Fig.~\ref{fig:Transition_vs_r}. The first quench from $1/(k_Fa_i) = -1.5$ to $1/(k_Fa_f) = -0.7$ (red diamonds), which is located in the transition regime between phase II and phase III, clearly demonstrates that $\Delta_\infty / \overline{\Delta}_f$ is independent of the aspect ratio $r$ in the transition between phases II and III as discussed in Sec.~\ref{sec:persistent}.
	  
	  The two other quenches in Fig.~\ref{fig:Transition_vs_r} are taken from the transition region between phase I and phase II. For the second quench from $1/(k_Fa_i) = -1.1$ to $1/(k_Fa_f) = -1.7$ (blue dots) we find a jump in $\Delta / \overline{\Delta}_f$ at $r = 30$, which corresponds to a shift of the onset of the dynamical vanishing to larger quench strengths with increasing aspect ratio. The same holds true for the third quench from $1/(k_Fa_i) = -0.3$ to $1/(k_Fa_f) = -1.0$ (black squares). However, here the jump of $\Delta_\infty / \overline{\Delta}_f$ at $r=42$ is smaller and the shift as a function of the aspect ratio is less pronounced due to the larger effective interaction strength $1/(k_Fa_f) = -1.0$, which leads to an exponentially larger $E_\text{min}$. The observed behavior for both quenches is in full agreement with our previously published result of a shift in the transition quench strength to phase I with $1/E_\text{min}$ where we found that $E_\text{min} \sim r$ \cite{Hannibal2018Dynamical}. 
	  
	  In a last step, we will further investigate the chaotic signature of the dynamics in the obtained dynamical phase diagram. To this end, we will discuss the phase diagram of the first Lyapunov exponent $x_{Ly}^{(1)}$ which reflects the most unstable mode in the entire system. The analysis is carried out for a system with an aspect ratio of $r=20$, which is exemplary for all other aspect ratios.
	  
\begin{figure}[t]
		\includegraphics[width=1\columnwidth]{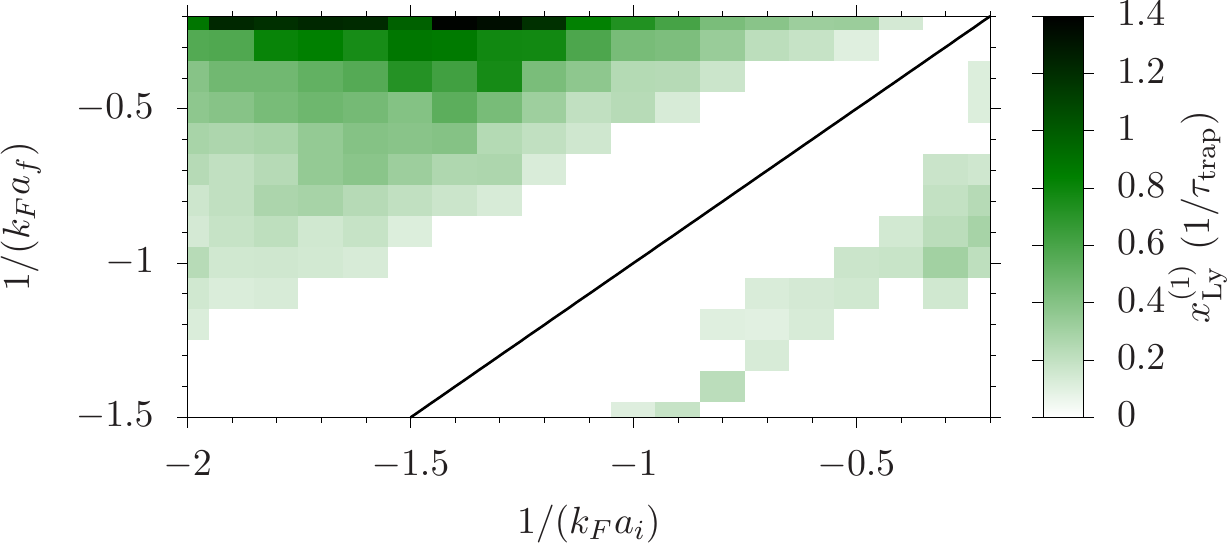}
		\caption{Phase diagram of the Lyapunov exponent $x_{Ly}^{(1)}$ for $r=20$ in units of $1/\tau_\text{trap}$. Due to our numerical implementation we cannot expect $x_{Ly}^{(1)}$ to vanish identically. Instead, we see a very slow increase of ln$(\left\|z'-z\right\| / \left\| z \right\|)$ even for small quenches which is not linear and saturates at values ln$(\left\|z'-z\right\| / \left\| z \right\|) < -15$. In these cases we set $x_{Ly}^{(1)} = 0$ since the characteristic of a chaotic system is not fulfilled.}
		\label{fig:chaos_phase}
	  \end{figure}

	  Figure~\ref{fig:chaos_phase} shows the phase diagram of the first Lyapunov exponent calculated by the numerical procedure presented in Sec.~\ref{sec:persistent}. We find $x_{Ly}^{(1)} = 0$ around the diagonal and in the lower right and lower left corner of the phase diagram. Furthermore, we observe two regions where the Lyapunov exponent takes a finite value: A stripe in the lower right part of the diagram and an extended region in the upper left corner. In the latter region the Lyapunov exponent increases with an increasing coupling strength in the final system $1/(k_Fa_f)$ while the dependence on the quench strength is nonmonotonic.
	  
	  Comparing the results for $x_{Ly}^{(1)}$ to the phase diagram of $\Delta_\infty / \overline{\Delta}$ for $r=20$ in Fig.~\ref{fig:phasedigram}~(a) yields a good agreement between the area of phase III and a finite and positive Lyapunov exponent. Hence, we confirm that the emergence of the dynamical phase III is indeed connected to a chaotic signature of the dynamics, which occurs if both the quench strength and the final coupling strength are sufficiently large.
	  
	  Furthermore, from comparing Fig.~\ref{fig:chaos_phase} to Fig.~\ref{fig:phasedigram}~(a) we obtain that the stripe-shaped region with a finite Lyapunov exponent corresponds to the transition regime between phase I and phase II. Here, we find a comparable situation as before: The quench strength, i.e., the distance to the diagonal (black line) is similar as for the onset of the chaos in phase III and the final coupling is still fairly large. As discussed above, this leads to a chaotic behavior of the system. Whereas, going deeper into phase I the coupling strength in the final system decreases and the dynamics is dominated by a dynamically vanishing order parameter, which is still nonlinear but does not show a chaotic signature due to the small coupling strength in the final system.
	  
	  Overall, we have obtained the dynamical phase diagram for an inhomogeneous Fermi gas after an interaction quench. We find three distinct phases, in agreement with the phases obtained in a homogeneous system in the thermodynamic limit \cite{Yuzbashyan2015Quantum}. We have analyzed the occurrence of each phase in dependence of the aspect ratio $r$. We demonstrated that phase III and phase II are unaffected by the aspect ratio for a constant system parameter $s$. Further, we have discussed the changes in the onset of phase I in dependence on the aspect ratio, which agrees with our previously published results \cite{Hannibal2018Dynamical}. Overall, we have found that the three-dimensional harmonic confinement alters the onset of the dynamical vanishing in phase I, leads to a breakdown of the damped oscillation in phase II, and introduces an amplitude modulation in phase III. Additionally, we have observed that the dynamics of phase III as well as in the transition region between phase I and phase II is connected to a chaotic behavior of the utilized BdG formalism.
	  
	  \section{conclusion}\label{sec:summary}
	  
	  In the current paper, we have analyzed the interaction quench dynamics of an ultracold Fermi gas quenched from the BCS regime to the BCS-BEC crossover regime. In the Higgs mode of the superfluid order parameter, we observe a nonlinear persistent oscillation with a small modulation of the amplitude. We have demonstrated that the frequency and the mean value of the oscillation are closely interconnected: Both are reduced simultaneously by the excitation of quasiparticles at the time of the quench which breaks Cooper pairs of the superfluid phase. These excitations are initially introduced at the time of the quench and hence they increase with increasing quench strength while they are independent of the aspect ratio for the case of a constant system parameter $s$. Furthermore, we have numerically demonstrated that the occurrence of the persistent oscillation is connected with a finite Lyapunov exponent of the dynamics which clearly reveals the chaotic signature of the dynamics in phase III in the BdG model.
	  
	  In Sec.~\ref{sec:phasediagramm} we have presented an interaction quench phase diagram of the Higgs mode for an ultracold Fermi gas confined in an harmonic trap. We observe three phases in accordance with the quench phase diagram in a homogeneous system \cite{Yuzbashyan2015Quantum}. For weak quenches of either direction we find damped oscillations dominated by linearly coupled oscillators which break down at $t_\text{trap}$. We observe that the location of phase II in the phase diagram around the diagonal is independent of the aspect ratio. For strong quenches ending in the deep BCS regime we find a dynamical vanishing of the order parameter. The onset of phase I is controlled by the ground-state properties of the system, i.e., by $E_\text{min}$, which is reflected in the dependence of the dynamical phase diagram on the aspect ratio. Finally, for large quenches towards the BCS-BEC crossover we find persistent nonlinear oscillations which are connected to a reduction of $\Delta_\infty / \overline{\Delta}_f$. The latter has been found to be independent of the aspect ratio. Therefore, the dynamical phase III as well as the transition between phase II and phase III is unaffected by a change of the aspect ratio as we have demonstrated in the dynamical phase diagram.
	  

%

\end{document}